\documentclass[twocolumn,superscriptaddress,prl,aps]{revtex4-1}
\usepackage{amssymb}
\usepackage{amsmath}
\usepackage{graphicx}
\usepackage{color}

\input{tcilatex}

\begin{document}

\title{Symmetry-Protected Scattering in Non-Hermitian Linear Systems}
\author{L. Jin}
\email{jinliang@nankai.edu.cn}
\author{Z. Song}
\affiliation{School of Physics, Nankai University, Tianjin 300071, China}

\begin{abstract}
Symmetry plays fundamental role in physics and the nature of symmetry
changes in non-Hermitian physics. Here the symmetry-protected scattering in
non-Hermitian linear systems is investigated by employing the discrete
symmetries that classify the random matrices. The even-parity symmetries
impose strict constraints on the scattering coefficients: the time-reversal (%
$C$ and $K$) symmetries protect the symmetric transmission or reflection;
the pseudo-Hermiticity ($Q$ symmetry) or the inversion ($P$) symmetry
protects the symmetric transmission and reflection. For the
inversion-combined time-reversal symmetries, the symmetric features on the
transmission and reflection interchange. The odd-parity symmetries including
the particle-hole symmetry, chiral symmetry, and sublattice symmetry cannot
ensure the scattering to be symmetric. These guiding principles are valid
for both Hermitian and non-Hermitian linear systems. Our findings provide
fundamental insights into symmetry and scattering ranging from condensed
matter physics to quantum physics and optics.
\end{abstract}

\maketitle

\textit{Introduction.---}Quantum transport and light scattering depend on
the properties of media \cite{Beenakker,Potton}. In optics, the Lorentz
reciprocity is fundamental due to the symmetric permittivity tensor; which
results in symmetric transmission when the input and output channels are
interchanged~\cite{Potton,AluNatPhoton}. Breaking reciprocity is important
for the light flow molding and the nonreciprocity plays a crucial role in
tailoring the light field. The optical isolator has a propagation direction
dependent transmission \cite{ZYu09,Qi12,Chang,LXQ15,LJinPRL,LDu20}. The
reciprocity breaks in the magneto-optical materials \cite%
{LBi,HRamezani12,Ganainy13,GanainyAPL13,HRamezani14} and the asymmetric
nonlinear optical structures \cite{Lepri,Kottos13,Sounas,YShi,Jing18}. In
comparison, the nonlinear systems are preferable for their integrability;
however, in addition to the requirement of high intensity, the desirable
light flow engineering is also comparably difficult. Alternatively, temporal
modulation of the propagation constant in the linear waveguides realizes
magnetic-free nonreciprocity through the synthesized magnetic flux \cite%
{Fang12a,Fang12b,Rechtsman13,Tzuang14,Li14,Longhi14,Goldman18,Cooper19,Ozawa19,Segev19,Szameit20,SFan20}%
, being advantageous for the scalable integrated devices in a wide range of
optical, radio, and audible frequencies \cite{AluNatPhoton}.

Recently, reciprocal and nonreciprocal anomalous scattering are demonstrated
in non-Hermitian systems \cite{Krasnok19}. The scattering dynamics closely
relates to the symmetries of the scattering center. The reciprocity still
holds in the parity-time-symmetric non-Hermitian metamaterials that
judiciously incorporate gain and loss~\cite{Jalas,Fan12,Yin13}. The
inversion symmetry guarantees the symmetric transmission and reflection \cite%
{Chong10,Wan2011,CPA,Baranov17,LonghiOL18,Trainiti19,Zhong20,Haque20}; the
time-reversal symmetry ensures the symmetric reflection \cite{Cannata,LJinSR}%
; and the parity-time symmetry protects the symmetric transmission \cite%
{LJinSR,Muga,LonghiJRA11,Longhi10,Stone11,Lin11,Feng11,Kottos,LGe12,Regensburger12, Feng13,Engheta13,Schomerus,Ahmed,Ambichl13,Savoia,Mostafazadeh,Wu14,Ramezani14,Fleury15, JLi15,YHuang15,Miri16,HZhao16,JHWu,LGe16,Wong16,LuPRL16,Ramezani16,JLi16,Ali16,JLi17,JHWuPRA17,LGe17,Schmelcher17,Gong17, Jin18,YLai18,YZhang18,Koutserimpas18,ShenPRM18,LeeOE18,Christensen18,Zhi18,Wu19,Muga19,Zhao19,SweeneyPRL19,Sweeney19,Novitsky20,Yang20}%
. Nevertheless, these conclusions are insufficient to fully capture the
symmetric properties of scattering and the role played by the symmetry for
an arbitrary linear system \cite{Rotter17}. More important, the nature of
symmetry changes in non-Hermitian physics \cite{KawabataNC19,KawabataPRX19}.
Now, the fundamental principles for the symmetry-protected scattering remain
concealed and are urgent to be settled as the rapid progresses in
non-Hermitian physics \cite%
{Konotop16,Kivshar,LFeng,Longhi,Ganainy,Alu,LYang,YFChen}.

In this Letter, we report the symmetry-protected scattering in non-Hermitian
linear systems and reveal the fundamental roles played by the symmetries. We
show that the internal symmetries $C,K,Q,P$ that classify the non-Hermitian
random matrices protect the symmetric transmission and/or reflection \cite%
{BL}. The non-Hermiticity helps breaking the symmetry protection and enables
a various of intriguing asymmetric scattering in the linear photonic
lattices, which has promising applications as optical diode, isolator, and
modulator. The scattering theory tackles problems including light
propagation in dissipative metamaterial, on-chip functional photonic device
design, and quantum transport manipulation in mesoscopic.

\textit{Symmetries.---}The non-Hermitian scattering center $H_{c}$ is
classified under the discrete symmetries \cite{BL,Diff}%
\begin{eqnarray}
C\text{ }sym. &:&H_{c}=\epsilon _{c}cH_{c}^{T}c^{-1},cc^{\ast }=\pm \mathbf{1%
}, \\
K\text{ }sym. &:&H_{c}=\epsilon _{\Bbbk }\Bbbk H_{c}^{\ast }\Bbbk
^{-1},\Bbbk \Bbbk ^{\ast }=\pm \mathbf{1}, \\
Q\text{ }sym. &:&H_{c}=\epsilon _{q}qH_{c}^{\dag }q^{-1},q^{2}=\mathbf{1}, \\
P\text{ }sym. &:&H_{c}=\epsilon _{p}pH_{c}p^{-1},p^{2}=\mathbf{1}.
\end{eqnarray}%
$H_{c}^{T}$, $H_{c}^{\ast }$, and $H_{c}^{\dagger }$ are the transpose,
complex conjugation, and Hermitian conjugation of $H_{c}$, respectively. $c$%
, $\Bbbk $, $q$, $p$ are unitary operators. The signs $\epsilon _{c,\Bbbk
,q,p}=\pm 1$ denote the parity of symmetries $C,K,Q,P$. For non-Hermitian
scattering center ($H_{c}\neq H_{c}^{\dagger }$), both the $C$ and $K$
symmetries relate to the time-reversal symmetry $\epsilon _{c,\Bbbk }=+1$
and the particle-hole symmetry $\epsilon _{c,\Bbbk }=-1$ \cite{KawabataPRX19}%
. The $Q$ symmetry is pseudo-Hermitian for $\epsilon _{q}=+1$ and
pseudo-anti-Hermitian for $\epsilon _{q}=-1$ (also referred to as the chiral
symmetry \cite{KawabataPRX19}). The $P$ symmetry with even-parity $\epsilon
_{p}=+1$ is the inversion symmetry if $p$ is the identity matrix rotated by $%
90$ degrees. The $P$ symmetry with the odd-parity $\epsilon _{p}=-1$ is the
sublattice symmetry. The eight symmetries form an $E8$ Abelian group \cite%
{E8}.

The even parity ($\epsilon _{c,\Bbbk ,q,p}=+1$) symmetries, including
time-reversal symmetry, pseudo-Hermiticity, and \textit{generalized}
inversion symmetry, can result in symmetric transmission and/or reflection;
the constraints imposed by the symmetries $C,K$ are that either the
transmission or the reflection is symmetric; both the symmetries $P,Q$ can
induce symmetric transmission and reflection. The symmetry-protected
transmission or reflection was observed in many experiments~\cite%
{Wong16,Wan2011,Feng13,BYan,Fleury14,Sliwa15,Regensburger12,Sliwa15,Fleury15}%
. In contrast, the odd-parity ($\epsilon _{c,\Bbbk ,q,p}=-1$) symmetries,
including particle-hole symmetry, chiral symmetry, and sublattice symmetry,
do not ensure the transmission or reflection to be symmetric because they
cannot impose any symmetric constraint on the scattering coefficients.

\textit{Scattering formalism.---}We consider a general multi-port linear
scattering center to elucidate the symmetry protection. In Fig.~\ref{fig1},
the scattering center is a time-independent $N$-site network (shaded in
orange). The schematic models physical systems including the coupled
resonators \cite{Ramezani14,LJinPRL,LYou,HCWu20}, coupled waveguides \cite%
{Wan2011,Feng11,Feng13,Konotop,Wong16}, and optical lattices \cite%
{HOtt,BYan,Wu14,RamezaniPRL18}. The solid circles stand for the resonators,
the waveguides, and the sites of the optical lattice. The solid lines
represent the couplings. The leads are uniform lattice chains with the
coupling strength $J$. For identical lead couplings, the scattering features
are fully determined by the properties of the scattering center. The $j$-th
lead is connected to the scattering center site $j$ at the coupling strength 
$g_{j}$. The arrows illustrate the scattering for the individual incidence
in the $m$-th and $n$-th leads, respectively. The outgoing waves in red
(green) are the reflections (transmissions). For more than one inputs, the
scattering wavefunction is a \textit{superposition} of wavefunctions of
separately injecting each individual input; thus, the scattering properties
are fully captured by the scattering of the individual input.

\begin{figure}[tb]
\includegraphics[ bb=35 360 560 560, width=8.8 cm, clip]{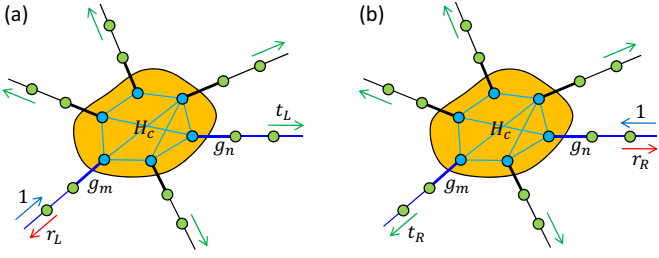} 
\caption{Schematic of a multi-port discrete scattering system. The orange area indicates an $N$-site scattering center $H_c$. The connection coupling
between the lead $j$ and the scattering center site $j$ is denoted as $g_j$ ($j\in \left[ 1,N\right]$). (a) Forward
incidence in the lead-$m$. (b) Backward incidence in the lead-$n$.} \label%
{fig1}
\end{figure}

In the coupled mode theory \cite{Haus,SFan03,Joannopoulos}, the equation of
motion for the monochromatic light field amplitude $\phi _{l,j}^{k}\left(
s\right) =\psi _{l,j}^{k}\left( s\right) e^{-i\omega t}$ in the $j$-th lead
is 
\begin{equation}
i\dot{\phi}_{l,j}^{k}(s)=\omega _{0}\phi _{l,j}^{k}(s)+J\phi
_{l,j}^{k}(s-1)+J\phi _{l,j}^{k}(s+1),  \label{1}
\end{equation}%
for the site $\left\vert s\right\vert \geq 1$ \cite{Ramezani14,Estep14}. The
dispersion relation supported by the leads is $\omega =\omega _{0}+2J\cos k$
for the incident momentum $k$ \cite{LJinPRL,Jin18}, obtained from the
steady-state solution of the light field amplitudes \cite{Jin10a,Jin10b}.
The resonant incidence has frequency $\omega _{0}$. The equations of motion
for the light field in the scattering center are 
\begin{equation}
i\left( 
\begin{array}{c}
\dot{\phi}_{c,1}^{k} \\ 
\vdots \\ 
\dot{\phi}_{c,N}^{k}%
\end{array}%
\right) =\left( \omega _{0}\mathbf{1+}H_{c}\right) \left( 
\begin{array}{c}
\phi _{c,1}^{k} \\ 
\vdots \\ 
\phi _{c,N}^{k}%
\end{array}%
\right) +\left( 
\begin{array}{c}
g_{1}\phi _{l,1}^{k}(1) \\ 
\vdots \\ 
g_{N}\phi _{l,N}^{k}(1)%
\end{array}%
\right) ,  \label{2}
\end{equation}%
where the $N\times N$ matrix $H_{c}$ characterizes the scattering center and 
$\mathbf{1}$ is the $N\times N$ identity matrix. $\phi _{c,j}^{k}$ is the
light field amplitude of the scattering center site $j$. $\phi _{l,j}^{k}(1)$
is the light field amplitude of the connection site on the lead $j$. $g_{j}$
is chosen $J$ or $0$ without loss of generality to indicate the presence or
absence of the lead $j$. For other $g_{j}$, the connection sites are counted
as part of the scattering center and the connection couplings remain $J$.
Setting $\phi _{c,j}^{k}=\psi _{c,j}^{k}e^{-i\omega t}$, we have $d\psi
_{c,j}^{k}/dt=0$ at the steady-state; and the equations of motion reduce to 
\begin{equation}
\omega \left( 
\begin{array}{c}
\psi _{c,1}^{k} \\ 
\vdots \\ 
\psi _{c,N}^{k}%
\end{array}%
\right) =\left( \omega _{0}\mathbf{1+}H_{c}\right) \left( 
\begin{array}{c}
\psi _{c,1}^{k} \\ 
\vdots \\ 
\psi _{c,N}^{k}%
\end{array}%
\right) +\left( 
\begin{array}{c}
g_{1}\psi _{l,1}^{k}(1) \\ 
\vdots \\ 
g_{N}\psi _{l,N}^{k}(1)%
\end{array}%
\right) .  \label{3}
\end{equation}

In the multi-port scattering center, we consider the scattering properties
of input and output in the leads $m$ and $n$. The steady-state equations of
motion for the multi-port scattering system are equivalent to that for a
two-port scattering system with leads $m$ and $n$. Each of the other lead $j$
($j\neq m,n$) effectively reduces into an additional on-site self-energy
term of the scattering center site $j$ in the equivalent scattering center $%
H_{c}^{\prime }$ \cite{Jin10b}. Notably, the wavefunction in the additional
lead $j$ ($j\neq m,n$) is outgoing wave $\psi _{l,j}^{k}\left( s\right)
=t_{j}e^{iks}$, and the wavefunction continuity yields $\psi
_{l,j}^{k}\left( 0\right) =\psi _{c,j}^{k}$. Thus, we have the relation $%
g_{j}\psi _{l,j}^{k}\left( 1\right) =g_{j}^{2}J^{-1}e^{ik}\psi
_{l,j}^{k}\left( 0\right) =g_{j}^{2}J^{-1}e^{ik}\psi _{c,j}^{k}$;
consequently, the second term $g_{j}\psi _{l,j}^{k}\left( 1\right) $ in Eq.~(%
\ref{3}) results in an extra self-energy $g_{j}^{2}J^{-1}e^{ik}$ for the
scattering center site $j$ in the equations of motion \cite{LJinJPA}, and
the multi-port scattering center is effectively characterized by the
two-port scattering center $H_{c}^{\prime }=H_{c}+J^{-1}e^{ik}\mathrm{diag}%
(\cdots ,g_{m-1}^{2},0,g_{m+1}^{2},\cdots ,g_{n-1}^{2},0,g_{n+1}^{2},\cdots
) $ with additional on-site complex self-energies except for the scattering
center sites $m$ and $n$. Therefore, the scattering properties of the
multi-port scattering center $H_{c}$ are completely determined from
analyzing the two-port scattering center $H_{c}^{\prime }$, and we focus on
investigating the scattering properties of the two-port scattering center.

We take $g_{m}=g_{n}=J$ and $g_{j}=0$ ($j\neq m,n$). From Eq.~(\ref{3}), the
wavefunctions for the scattering center sites $m$ and $n$ satisfy%
\begin{eqnarray}
\psi _{c,m}^{k} &=&-\Delta _{mm}^{-1}J\psi _{l,m}^{k}(-1)-\Delta
_{mn}^{-1}J\psi _{l,n}^{k}(1),  \label{LeadL} \\
\psi _{c,n}^{k} &=&-\Delta _{nm}^{-1}J\psi _{l,m}^{k}(-1)-\Delta
_{nn}^{-1}J\psi _{l,n}^{k}(1),  \label{LeadR}
\end{eqnarray}%
where $\Delta _{mn}^{-1}$ is the element of the $m$-th row and $n$-th column
of the inverse matrix of $\Delta =H_{c}-\left( 2J\cos k\right) \mathbf{1}$~%
\cite{NoteInverseMatrix}. For the multi-port case, just replace $H_{c}$ with 
$H_{c}^{\prime }$ in $\Delta $.

We index $-1$ to $-\infty $ for sites of the left lead (lead $m$) and index $%
1$ to $+\infty $ for sites of the right lead (lead $n$). The stationary
states are the superpositions of incoming and outgoing waves \cite{Ambichl13}%
. The wavefunctions for the forward incidence $\psi _{L}^{k}(s)$ and
backward incidence $\psi _{R}^{k}(s)$ are two linearly independent solutions%
\begin{eqnarray}
\psi _{L}^{k}(s) &=&e^{iks}+r_{L}e^{-iks},(s<0);t_{L}e^{iks},(s>0),
\label{4} \\
\psi _{R}^{k}(s) &=&t_{R}e^{-iks},(s<0);e^{-iks}+r_{R}e^{iks}(s>0).
\label{5}
\end{eqnarray}%
The wavefunction continuity $\psi _{c,m}^{k}=\psi _{l,m}^{k}(0)$, $\psi
_{c,n}^{k}=\psi _{l,n}^{k}(0)$ yields $\psi _{c,m}^{k}=1+r_{L}$, $\psi
_{c,n}^{k}=t_{L}$ for the forward incidence; from Eq.~(\ref{4}), we have $%
\psi _{l,m}^{k}(-1)=e^{-ik}+r_{L}e^{ik}$, $\psi _{l,n}^{k}(1)=t_{L}e^{ik}$.
Substituting these wavefunctions into Eqs.~(\ref{LeadL}) and~(\ref{LeadR}),
we obtain $t_{L}$ and $r_{L}$. For the backward incidence, we have $\psi
_{c,m}^{k}=t_{R}$, $\psi _{c,n}^{k}=1+r_{R}$; from Eq. (\ref{5}), we have $%
\psi _{l,m}^{k}(-1)=t_{R}e^{ik}$, $\psi _{l,n}^{k}(1)=e^{-ik}+r_{R}e^{ik}$.
Substituting these wavefunctions into Eqs.~(\ref{LeadL}) and~(\ref{LeadR}),
we obtain $t_{R}$ and $r_{R}$. The scattering coefficients are%
\begin{equation}
\begin{array}{c}
t_{L}=\frac{\Delta _{nm}^{-1}J^{-1}(e^{ik}-e^{-ik})}{(J^{-1}+\Delta
_{mm}^{-1}e^{ik})(J^{-1}+\Delta _{nn}^{-1}e^{ik})-\Delta _{mn}^{-1}\Delta
_{nm}^{-1}e^{2ik}}, \\ 
r_{L}=\frac{\Delta _{mn}^{-1}\Delta _{nm}^{-1}-(J^{-1}e^{ik}+\Delta
_{mm}^{-1})(J^{-1}e^{-ik}+\Delta _{nn}^{-1})}{(J^{-1}+\Delta
_{mm}^{-1}e^{ik})(J^{-1}+\Delta _{nn}^{-1}e^{ik})-\Delta _{mn}^{-1}\Delta
_{nm}^{-1}e^{2ik}}, \\ 
t_{R}=\frac{\Delta _{mn}^{-1}J^{-1}(e^{ik}-e^{-ik})}{(J^{-1}+\Delta
_{mm}^{-1}e^{ik})(J^{-1}+\Delta _{nn}^{-1}e^{ik})-\Delta _{mn}^{-1}\Delta
_{nm}^{-1}e^{2ik}}, \\ 
r_{R}=\frac{\Delta _{mn}^{-1}\Delta _{nm}^{-1}-(J^{-1}e^{ik}+\Delta
_{nn}^{-1})(J^{-1}e^{-ik}+\Delta _{mm}^{-1})}{(J^{-1}+\Delta
_{mm}^{-1}e^{ik})(J^{-1}+\Delta _{nn}^{-1}e^{ik})-\Delta _{mn}^{-1}\Delta
_{nm}^{-1}e^{2ik}}.%
\end{array}
\label{6}
\end{equation}%
The symmetric transmission is%
\begin{equation}
t_{L}=t_{R}\text{ for }\Delta _{mn}^{-1}=\Delta _{nm}^{-1};\left\vert
t_{L}\right\vert =\left\vert t_{R}\right\vert \text{ for }|\Delta
_{mn}^{-1}|=|\Delta _{nm}^{-1}|.  \label{T}
\end{equation}%
The symmetric reflection is%
\begin{equation}
\begin{array}{l}
r_{L}=r_{R}\text{ for }\Delta _{mm}^{-1}=\Delta _{nn}^{-1}; \\ 
\left\vert r_{L}\right\vert =\left\vert r_{R}\right\vert \text{ for real }%
\Delta _{mm}^{-1},\Delta _{nn}^{-1},\Delta _{mn}^{-1}\Delta _{nm}^{-1}.%
\end{array}
\label{R}
\end{equation}%
The scattering properties of each pair of input-output leads are
straightforwardly obtained in this manner. The symmetries of the scattering
center $H_{c}$, imposing restrict constraints on the scattering
coefficients, are \textit{essential} to understand the symmetric scattering
dynamics.

\textit{Symmetry protection.---}The symmetry-protected scattering properties
are closely related to the spatial structure of the scattering center and
rely on two ways of mapping%
\begin{equation}
\mathbf{1}:U_{\mathbf{1}}\left\vert m(n)\right\rangle _{c}\rightarrow
\left\vert m(n)\right\rangle _{c};\mathcal{I}:U_{\mathcal{I}}\left\vert
m(n)\right\rangle _{c}\rightarrow \left\vert n(m)\right\rangle _{c},
\label{MappingRelation}
\end{equation}%
where $U_{\mathbf{1},\mathcal{I}}=c,\Bbbk ,q,p$; $\left\vert m\right\rangle
_{c}$ and $\left\vert n\right\rangle _{c}$ denote the two connection sites
of the scattering center that are connected with the leads $m$ and $n$,
respectively. The mapping manners reclassify eight even-parity symmetries%
\begin{equation}
C_{\mathbf{1}},C_{\mathcal{I}};K_{\mathbf{1}},K_{\mathcal{I}};Q_{\mathbf{1}%
},Q_{\mathcal{I}};P_{\mathbf{1}},P_{\mathcal{I}}.
\end{equation}%
The subscripts indicate the mapping manners. The $P_{\mathbf{1}}$ symmetry
is trivial. The $P_{\mathcal{I}}$ symmetry is a \textit{generalized}
inversion symmetry and leads to symmetric transmission and reflection \cite%
{Chong10,Wan2011,CPA,Baranov17,LonghiOL18,Trainiti19,Zhong20,Haque20}. If
the scattering center \textit{only} has the $C_{\mathbf{1}}$ or $K_{\mathcal{%
I}}$ symmetry, the transmission is symmetric; but the reflection is
asymmetric due to the lack of symmetry protection unless Eq.~(\ref{R}) is
satisfied. Similarly, if the scattering center \textit{only} has the $C_{%
\mathcal{I}}$ or $K_{\mathbf{1}}$ symmetry, the reflection is symmetric, but
the transmission is asymmetric unless Eq.~(\ref{T}) is satisfied. Table~\ref%
{TableI} lists the constraints for the corresponding symmetries
(Supplemental Material A). For the scattering center without the $C_{\mathbf{%
1}},K_{\mathcal{I}},C_{\mathcal{I}},K_{\mathbf{1}},P_{\mathcal{I}},Q_{%
\mathbf{1}}$ symmetries, both the transmission and reflection are asymmetric
unless Eq.~(\ref{T}) or (\ref{R}) is satisfied. The symmetry protection
still valid even though the scattering coefficients diverge at the spectral
singularity in the anomalous scattering \cite{Krasnok19,PWang16}, where
lasing occurs as the time-reversal process of perfect absorbing \cite%
{Ramezani14,LJinPRL}. $H_{c}=H_{c}^{T}$ belongs to the $C_{\mathbf{1}}$
symmetry and characterizes the Lorentz reciprocity in optics \cite%
{Potton,Jalas,AluNatPhoton,LJinPRA18}. The scattering center $%
H_{c}=[1,e^{-i\phi };e^{i\phi },i]$ has the $C_{\mathbf{1}}$ symmetry, where
the unitary operator is $c=\mathrm{diag}(1,e^{2i\phi })$ and the mapping
between the connection sites is $c[\left\vert m\right\rangle ,\left\vert
n\right\rangle ]^{T}=[\left\vert m\right\rangle ,e^{2i\phi }\left\vert
n\right\rangle ]^{T}$; consequently, $t_{L}=e^{2i\phi }t_{R}$. The $C_{%
\mathcal{I}}$ symmetry is the combined $P_{\mathcal{I}}C_{\mathbf{1}}$
symmetry and leads to the symmetric reflection $r_{L}=r_{R}$ \cite%
{XZZhang13,CLi17}. The $K$ symmetry has the complex conjugation operation
and relates to the time-reversal symmetry \cite{KawabataPRX19}. The $K_{%
\mathbf{1}}$ symmetry results in the symmetric reflection $\left\vert
r_{L}\right\vert =\left\vert r_{R}\right\vert $ \cite{Cannata,LJinSR}. The $%
K_{\mathcal{I}}$ symmetry is the combined $P_{\mathcal{I}}K_{\mathbf{1}}$
symmetry and results in the symmetric transmission $\left\vert
t_{L}\right\vert =\left\vert t_{R}\right\vert $; the parity-time symmetry
belongs to the $K_{\mathcal{I}}$ symmetry \cite%
{LJinSR,Muga,LonghiJRA11,Longhi10,Stone11,Lin11,Feng11,Kottos,LGe12,Regensburger12,Feng13,Engheta13,Schomerus,Ahmed,Ambichl13,Savoia,Mostafazadeh,Wu14,Ramezani14,Fleury15,JLi15,YHuang15,Miri16,HZhao16,JHWu,LGe16,Wong16,LuPRL16,Ramezani16,JLi16, Ali16,JLi17,JHWuPRA17,LGe17,Schmelcher17,Gong17,Jin18,YLai18,YZhang18,Koutserimpas18,ShenPRM18,LeeOE18,Christensen18,Zhi18,Wu19,Muga19,Zhao19,SweeneyPRL19,Sweeney19,Novitsky20,Yang20}%
. The $Q_{\mathbf{1}}$ symmetry is a combined $C_{\mathbf{1}}K_{\mathbf{1}}$
or $C_{\mathcal{I}}K_{\mathcal{I}}$ symmetry. For example, $%
H_{c}=[0,i-1;i+1,1]$ has the $Q_{\mathbf{1}}$ symmetry with $q=\sigma _{z}$;
the Hermitian scattering centers possess the $Q_{\mathbf{1}}$ symmetry with $%
q$ being the identity matrix. The transmission and reflection are both
symmetric under the $Q_{\mathbf{1}}$ symmetry protection \cite{LJin12}. In
contrast, the combined $C_{\mathbf{1}}K_{\mathcal{I}}$ or $C_{\mathcal{I}}K_{%
\mathbf{1}}$ symmetry: the $Q_{\mathcal{I}}$ symmetry, imposes no symmetric
constraint on the scattering coefficients. For example, both the
transmission and reflection are asymmetric for the $Q_{\mathcal{I}}$
symmetric non-Hermitian scattering center $H_{c}=[i\gamma ,Je^{-\varphi
};Je^{\varphi },-i\gamma ]$, the corresponding unitary operator is $q=\sigma
_{x}$; the cooperation between asymmetric coupling~\cite%
{Schomerus15,LonghiPRB15,KhajavikhanOE18,LJinPRB19}, gain, and loss destroys
the symmetric transmission and reflection.

\begin{table}[tb]
\caption{Symmetry-protected constraint on the transmission and reflection
for each individual symmetry.} \label{TableI}%
\begin{tabular}{|c|c|c|c|c|}
\hline
\textrm{Symmetry} & $C_{\mathbf{1}},K_{\mathcal{I}}$ & $C_{\mathcal{I}},K_{%
\mathbf{1}}$ & $Q_{\mathbf{1}},P_{\mathcal{I}}$ & $Q_{\mathcal{I}},P_{%
\mathbf{1}}$ \\ \hline
\textrm{Constraint} & $\left\vert t_{L}\right\vert =\left\vert
t_{R}\right\vert $ & $\left\vert r_{L}\right\vert =\left\vert
r_{R}\right\vert $ & \textrm{Both} & \textrm{None} \\ \hline
\end{tabular}
\end{table}

\begin{figure}[tb]
\includegraphics[ bb=10 0 575 170, width=8.8 cm, clip]{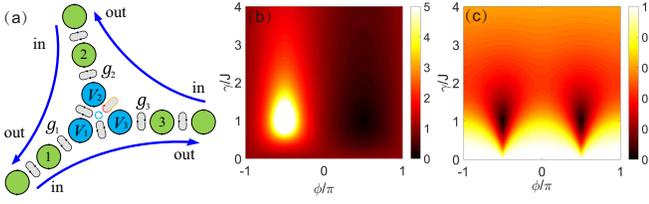} 
\caption{(a) Schematic of the three-coupled-resonator scattering center. (b) $|t_R|/|t_L|$ for both
$\left\{i\protect\gamma, -i\protect\gamma ,0 \right\}$ and $\left\{0,
-i\protect\gamma ,0 \right\}$. $|t_R|/|t_L|$ diverges at
$\protect\phi=-\protect\pi/2$, $\protect\gamma/J=1$ and is cut to $5$. (c)
$|r_R|/|r_L|$ for $\left\{i\protect\gamma, -i\protect\gamma ,0 \right\}$. (b, c) are for $g_{1}=g_{3}=J$ and $g_2=0$ (Supplemental Material B). The incidence has resonant frequency $\protect\omega_0$.}
\label{fig2}
\end{figure}

\textit{Breaking reciprocity.}---Under the guidance of symmetry protection,
we exemplify asymmetric light scattering in a three-coupled-resonator
scattering center. In Fig.~\ref{fig2}(a), the primary resonators
(round-shape) are effectively coupled through the link resonators
(stadium-shape). The Peierls phase factor $e^{\pm i\phi }$ presents in one
of the couplings between the central three resonators~\cite%
{Peterson18,LJinPRA93,LJinPRA97}, where photons tunneling in the forward and
backward directions experience different path lengths as indicated by the
orange and red arrows in the link resonator \cite{Hafezi13,Hafezi14}. In the
equations of motion [Eq.~(\ref{2})], the scattering center is 
\begin{equation}
H_{c}=\left( 
\begin{array}{ccc}
V_{1} & J & J \\ 
J & V_{2} & Je^{-i\phi } \\ 
J & Je^{i\phi } & V_{3}%
\end{array}%
\right) ,
\end{equation}%
where $\omega _{0}+$\textrm{Re}$(V_{j})$ and \textrm{Im}$(V_{j})$ are the
resonant frequency and gain/loss for the resonator $j=1,2,3$, respectively.

We take $g_{1}=g_{3}=J$, $g_{2}=0$ and discuss the symmetry protection.
Without the gain and loss, the $Q_{\mathbf{1}}$ symmetry ensures the
symmetric transmission and reflection. Without the nonreciprocal coupling,
the $C_{\mathbf{1}}$ symmetry ensures the symmetric transmission. The
interplay between the gain/loss and nonreciprocal coupling generates the
asymmetric transmission. $H_{c}$ with complex $V_{1}=V_{3}^{\ast }$ and real 
$V_{2}\neq 0$ only has the $K_{\mathcal{I}}$ symmetry,%
\begin{equation}
\Bbbk H_{c}^{\ast }\Bbbk ^{-1}=\left( 
\begin{array}{ccc}
V_{3}^{\ast } & J & J \\ 
J & V_{2}^{\ast } & Je^{-i\phi } \\ 
J & Je^{i\phi } & V_{1}^{\ast }%
\end{array}%
\right) ,\Bbbk =\left( 
\begin{array}{ccc}
0 & 0 & 1 \\ 
0 & e^{-i\phi } & 0 \\ 
1 & 0 & 0%
\end{array}%
\right) .
\end{equation}%
Thus, the transmission is symmetric, but the reflection is asymmetric \cite%
{Ramezani14}. $H_{c}$ with $V_{1}=V_{3}$ and $V_{2}\neq V_{2}^{\ast }$ only
has the $C_{\mathcal{I}}$ symmetry \cite{LJin16},%
\begin{equation}
cH_{c}^{T}c^{-1}=\left( 
\begin{array}{ccc}
V_{3} & J & J \\ 
J & V_{2} & Je^{-i\phi } \\ 
J & Je^{i\phi } & V_{1}%
\end{array}%
\right) ,c=\left( 
\begin{array}{ccc}
0 & 0 & 1 \\ 
0 & e^{-i\phi } & 0 \\ 
1 & 0 & 0%
\end{array}%
\right) .
\end{equation}

A single loss center $\left\{ V_{1},V_{2},V_{3}\right\} =\left\{ 0,-i\gamma
,0\right\} $ has asymmetric transmission, but symmetric reflection; an ideal
optical isolator with $S$-matrix $S=[r_{L},t_{R};t_{L},r_{R}]=(-i\sigma
_{x}\pm \sigma _{y})/2$ is generated at $J=\gamma $ and $\phi =\mp \pi /2$
for resonant incidence $k=-\pi /2$ as indicated in Fig.~\ref{fig2}(b). In
contrast, $\left\vert V_{1}\right\vert \neq \left\vert V_{3}\right\vert $
breaks all the symmetries of $H_{c}$ and both the transmission and
reflection are asymmetric.

\begin{figure}[tb]
\includegraphics[ bb=0 0 560 145, width=8.8 cm, clip]{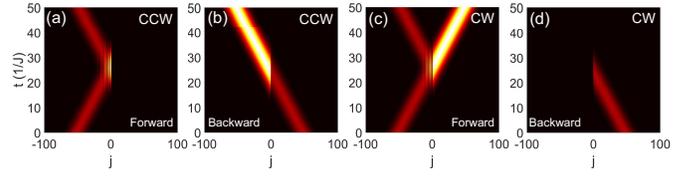} 
\caption{Asymmetric scattering dynamics for Figs.~\ref{fig2}(b, c) at $\protect\gamma =J$. (a) Forward and (b) backward incidences
for $\protect\phi =-\protect\pi /2$. The counterclockwise (CCW) mode and the
clockwise (CW) mode of the resonator experience opposite magnetic fluxes.
(c) Forward and (d) backward incidences for $\protect\phi =\protect\pi /2$.} %
\label{fig3}
\end{figure}

The striking asymmetric scattering for $\left\{ V_{1},V_{2},V_{3}\right\}
=\left\{ i\gamma ,-i\gamma ,0\right\} $ in Figs.~\ref{fig2}(b) and \ref{fig2}%
(c) indicates a chiral perfect absorption \cite{SweeneyPRL19} that
unidirectional incidence is completely absorbed for the clockwise mode \cite%
{Longhi15,Ramezani16,LJin16}. A unidirectional transmissionless $t_{L}=0$, $%
r_{L}=1$; $t_{R}=-2i$, $r_{R}=0$ occurs at $\gamma =J$, $\phi =-\pi /2$
[Figs.~\ref{fig3}(a) and~\ref{fig3}(b)] and a unidirectional absorption $%
t_{L}=-2i$, $r_{L}=1$; $t_{R}=0$, $r_{R}=0$ occurs at $\gamma =J$, $\phi
=\pi /2$ [Figs.~\ref{fig3}(c) and~\ref{fig3}(d)] for resonant incidence $%
k=-\pi /2$.

For the three-port scattering, the transmission in the lead $3$ ($2$) for
incidence in the lead $1$ is straightforwardly obtained from Eq.~(\ref{6})
by taking $m=1$, $n=3$($2$) and replacing $V_{2(3)}$ with $V_{2(3)}^{\prime
}=V_{2(3)}+Je^{ik}$ in $H_{c}$ \cite{LJinJPA}. At $\left\{
V_{1},V_{2},V_{3}\right\} =\left\{ 0,0,0\right\} $, the input resonantly
outgoes from one of the adjacent leads as indicated by the blue arrows at $%
\phi =\pi /2$ and inversely at $\phi =-\pi /2$ for the resonant incidence,
and functions as a circulator~\cite{Fleury14,Estep14,Sliwa15} with symmetric
zero reflection protected by the $C_{\mathcal{I}}$ symmetry because $%
V_{2}^{\prime }\neq (V_{2}^{\prime })^{\ast }$.

Without the gain and loss, the non-Hermitian dissipative coupling \cite%
{LYou,antiPT,Konotop,CWQiu,SWDu,HCWu20,CHHu20} associated with nonreciprocal
coupling can also break the symmetry protection and generate asymmetric
transmission and reflection in the three-coupled-resonator scattering center 
\begin{equation}
H_{c}=\left( 
\begin{array}{ccc}
0 & -i\kappa & J \\ 
-i\kappa & 0 & Je^{-i\phi } \\ 
J & Je^{i\phi } & 0%
\end{array}%
\right) .
\end{equation}%
More details are provided in Supplemental Material C.

\textit{Discussion.---}For the two-port linear scattering center, the
effective complex self-energy is absent. Thus, the non-Hermiticity is
required to realize asymmetric transmission because Hermitian systems are $%
Q_{\mathbf{1}}$ symmetry protected. The $C_{\mathbf{1}},K_{\mathcal{I}},P_{%
\mathcal{I}},Q_{\mathbf{1}}$ symmetries should be absent to break symmetric
transmission. In addition to the non-Hermiticity required to break the
pseudo-Hermiticity ($Q_{\mathbf{1}}$ symmetry), the nonreciprocal coupling
is required to break the $C_{\mathbf{1}}$ symmetry protection. The simplest
example is a two-site center with asymmetric coupling strengths $%
H_{c}=[0,Je^{-\varphi };Je^{\varphi },0]$ \cite{XZZhang13,CLi17}; the
three-coupled-resonator scattering center with $\left\{ 0,-i\gamma
,0\right\} $ is another example, and other examples include systems studied
in Refs. \cite%
{LXQ15,Longhi15,LJin16,LJinPRL,BYan,YFChenAPL11,YFChenPRL11,YFChenPLA12}.

The $C_{\mathcal{I}},K_{\mathbf{1}},P_{\mathcal{I}},Q_{\mathbf{1}}$
symmetries should be absent to break symmetric reflection. Provided that the
resonator gain and/or loss are not balanced, all these four symmetries are
absent. Thus, the asymmetric reflection ubiquitously presents in the
intriguing scattering phenomena, including the unidirectional reflectionless 
\cite{Fleury15,Feng13,Huang17,Regensburger12,HZhao16}, unidirectional lasing 
\cite{Ramezani14,LJinPRL}, coherent perfect absorber laser \cite%
{Longhi10,Stone11,Wong16,ZHou19}, chiral absorber \cite{SweeneyPRL19}, and
chiral metamaterials \cite{Droulias19,Katsantonis20}.

The situation $\left\vert V_{1}\right\vert \neq \left\vert V_{3}\right\vert $
in the three-coupled-resonator (Fig.~\ref{fig3}) and the systems studied in
Refs. \cite{LJinPRL,LJinNJP17} exemplify the asymmetric transmission and
reflection without the protection of all the six symmetries $C_{\mathbf{1}%
},C_{\mathcal{I}},K_{\mathbf{1}},K_{\mathcal{I}},P_{\mathcal{I}},Q_{\mathbf{1%
}}$. Properly incorporating nonreciprocal coupling, asymmetric coupling,
dissipative coupling, gain, and loss generate asymmetric transmission and
reflection.

For the multi-port scattering center, the effective scattering center $%
H_{c}^{\prime }$ may only possess the $C_{\mathbf{1}},C_{\mathcal{I}}$ or $%
P_{\mathcal{I}}$ symmetry due to the momentum dependent self-energy in $%
H_{c}^{\prime }$. Thus, asymmetric scattering behavior easily occurs in the
multi-port scattering center. If the leads are symmetrically coupled to the
scattering center $H_{c}$, the transmission and (or) reflection of the
multi-port scattering center are still symmetric under the $P_{\mathcal{I}}$
($C_{\mathbf{1}}$ or $C_{\mathcal{I}}$) symmetry protection. Notably, the
scattering properties of $H_{c}^{\prime }$ may be $K_{\mathbf{1}},K_{%
\mathcal{I}}$ or $Q_{\mathbf{1}}$ symmetry-protected at certain momentum.

The symmetries with opposite parities, various symmetry types, and different
ways of mapping may coexist in the scattering center. The constraints
imposed by the symmetries on the scattering coefficients coexist and affect
simultaneously. The symmetry protection provides fundamental guiding
principles for manipulating quantum transport in mesoscopic and tailoring
the light flow in the integrated photonics.

\textit{Conclusion.---}We unveil the roles played by the symmetry for the
scattering in non-Hermitian linear systems. The time-reversal symmetry,
pseudo-Hermiticity (including Hermiticity), and generalized inversion
symmetry protect the symmetric transmission and/or reflection (Table \ref%
{TableI}); however, the particle-hole symmetry, chiral symmetry, and
sublattice symmetry do not. These provide fundamental guiding principles for
the light scattering in both Hermitian and non-Hermitian systems. Our
findings are valid in the quantum systems and pave the way for further
investigations on the transport in non-Hermitian physics.

\acknowledgments We acknowledge the support of National Natural Science
Foundation of China (Grants No.~11975128 and No.~11874225). L.J.
acknowledges H. C. Wu for discussions and S. M. Zhang for useful comments on
a first draft of this paper.

\clearpage

\newpage 
\begin{widetext}

\section*{Supplemental Material for \textquotedblleft Symmetry-Protected
Scattering in Non-Hermitian Linear Systems"}

\begin{center}
L. Jin$^{1,\ast }$ and Z. Song$^{1}$\\[2pt]
\textit{$^{1}$School of Physics, Nankai University, Tianjin 300071, China}
\end{center}
\subsection*{A: Proof of the constraints on the scattering coefficients for
discrete symmetries}

\emph{Symmetric scattering coefficients}\textit{.---}The transmission and
reflection coefficients for the input-output in the leads $m$ and $n$ are%
\begin{eqnarray}
t_{L} &=&\frac{\Delta _{nm}^{-1}J^{-1}(e^{ik}-e^{-ik})}{(J^{-1}+\Delta
_{mm}^{-1}e^{ik})(J^{-1}+\Delta _{nn}^{-1}e^{ik})-\Delta _{mn}^{-1}\Delta
_{nm}^{-1}e^{2ik}},  \label{tL} \\
r_{L} &=&\frac{\Delta _{mn}^{-1}\Delta _{nm}^{-1}-(J^{-1}e^{ik}+\Delta
_{mm}^{-1})(J^{-1}e^{-ik}+\Delta _{nn}^{-1})}{(J^{-1}+\Delta
_{mm}^{-1}e^{ik})(J^{-1}+\Delta _{nn}^{-1}e^{ik})-\Delta _{mn}^{-1}\Delta
_{nm}^{-1}e^{2ik}}.  \label{rL} \\
t_{R} &=&\frac{\Delta _{mn}^{-1}J^{-1}(e^{ik}-e^{-ik})}{(J^{-1}+\Delta
_{mm}^{-1}e^{ik})(J^{-1}+\Delta _{nn}^{-1}e^{ik})-\Delta _{mn}^{-1}\Delta
_{nm}^{-1}e^{2ik}},  \label{tR} \\
r_{R} &=&\frac{\Delta _{mn}^{-1}\Delta _{nm}^{-1}-(J^{-1}e^{ik}+\Delta
_{nn}^{-1})(J^{-1}e^{-ik}+\Delta _{mm}^{-1})}{(J^{-1}+\Delta
_{mm}^{-1}e^{ik})(J^{-1}+\Delta _{nn}^{-1}e^{ik})-\Delta _{mn}^{-1}\Delta
_{nm}^{-1}e^{2ik}},  \label{rR}
\end{eqnarray}%
where $\Delta =H_{c}+\left( \omega _{0}-\omega \right) \mathbf{1=}%
H_{c}-\left( 2J\cos k\right) \mathbf{1}$, $\Delta ^{-1}$ indicates the
inverse of $\Delta $, and $\Delta _{mn}^{-1}$ indicates the element on the $%
m $-th row and the $n$-th column of $\Delta ^{-1}$.

It is obvious that
\begin{eqnarray}
t_{L} &=&t_{R}\text{ for }\Delta _{nm}^{-1}=\Delta _{mn}^{-1}, \\
r_{L} &=&r_{R}\text{ for }\Delta _{mm}^{-1}=\Delta _{nn}^{-1}.
\end{eqnarray}%
Alternatively,
\begin{equation}
\left\vert t_{L}\right\vert ^{2}=\left\vert t_{R}\right\vert ^{2}\text{ for }%
\left\vert \Delta _{nm}^{-1}\right\vert =\left\vert \Delta
_{mn}^{-1}\right\vert .
\end{equation}%
The difference between the numerators of $\left\vert r_{L}\right\vert ^{2}$
and $\left\vert r_{R}\right\vert ^{2}$ is given by%
\begin{eqnarray}
&&\left\vert \Delta _{mn}^{-1}\Delta _{nm}^{-1}-(J^{-1}e^{ik}+\Delta
_{mm}^{-1})(J^{-1}e^{-ik}+\Delta _{nn}^{-1})\right\vert ^{2}-\left\vert
\Delta _{mn}^{-1}\Delta _{nm}^{-1}-(J^{-1}e^{ik}+\Delta
_{nn}^{-1})(J^{-1}e^{-ik}+\Delta _{mm}^{-1})\right\vert ^{2} \\
&=&2J^{-2}i\sin \left( 2k\right) \left( \Delta _{nn}^{-1}\Delta
_{mm}^{-1\ast }-\Delta _{mm}^{-1}\Delta _{nn}^{-1\ast }\right)  \notag \\
&&+2J^{-1}i\sin k[\left( J^{-2}+\Delta _{mm}^{-1\ast }\Delta _{nn}^{-1\ast
}-\Delta _{mn}^{-1\ast }\Delta _{nm}^{-1\ast }\right) \left( \Delta
_{nn}^{-1}-\Delta _{mm}^{-1}\right) -\left( J^{-2}+\Delta _{mm}^{-1}\Delta
_{nn}^{-1}-\Delta _{mn}^{-1}\Delta _{nm}^{-1}\right) \left( \Delta
_{nn}^{-1\ast }-\Delta _{mm}^{-1\ast }\right) ].  \notag
\end{eqnarray}%
Thus, we have $\left\vert r_{L}\right\vert ^{2}=\left\vert r_{R}\right\vert
^{2}$ for real $\Delta _{mm}^{-1}$, $\Delta _{nn}^{-1}$, and $\Delta
_{mn}^{-1}\Delta _{nm}^{-1}$. These are the conclusions of Eqs.~(9) and (10)
in the main text. Notably, the accidental symmetric reflection $\left\vert
r_{L}\right\vert ^{2}=\left\vert r_{R}\right\vert ^{2}$ occurs if the
conditions $J^{-2}+\Delta _{mm}^{-1\ast }\Delta _{nn}^{-1\ast }-\Delta
_{mn}^{-1\ast }\Delta _{nm}^{-1\ast }=0$ and $\Delta _{nn}^{-1}\Delta
_{mm}^{-1\ast }=\Delta _{mm}^{-1}\Delta _{nn}^{-1\ast }$ are simultaneously
satisfied.

\emph{Symmetry-protected scattering}\textit{.---}The mapping relations are%
\begin{eqnarray}
\mathbf{1} &:&U_{\mathbf{1}}[\cdots ,\left\vert m\right\rangle _{c},\cdots
,\left\vert n\right\rangle _{c},\cdots ]^{T}=[\cdots ,\left\vert
m\right\rangle _{c},\cdots ,e^{i\alpha }\left\vert n\right\rangle
_{c},\cdots ]^{T}, \\
\mathcal{I} &:&U_{\mathcal{I}}[\cdots ,\left\vert m\right\rangle _{c},\cdots
,\left\vert n\right\rangle _{c},\cdots ]^{T}=[\cdots ,\left\vert
n\right\rangle _{c},\cdots ,e^{i\alpha }\left\vert m\right\rangle
_{c},\cdots ]^{T}.
\end{eqnarray}%
In the general situation, the phase factor $e^{i\alpha }$ is not necessarily
to be $1$. The four elements in the unitary operators that relate to the
mapping between the connection sites $\left\vert m\right\rangle _{c}$ and $%
\left\vert n\right\rangle _{c}$ are%
\begin{eqnarray}
\mathbf{1} &:&\left(
\begin{array}{cc}
U_{\mathbf{1,}mm} & U_{\mathbf{1,}nm} \\
U_{\mathbf{1,}mn} & U_{\mathbf{1,}nn}%
\end{array}%
\right) =\left(
\begin{array}{cc}
1 & 0 \\
0 & e^{i\alpha }%
\end{array}%
\right) , \\
\mathcal{I} &:&\left(
\begin{array}{cc}
U_{\mathcal{I}\mathbf{,}mm} & U_{\mathcal{I}\mathbf{,}nm} \\
U_{\mathcal{I}\mathbf{,}mn} & U_{\mathcal{I}\mathbf{,}nn}%
\end{array}%
\right) =\left(
\begin{array}{cc}
0 & 1 \\
e^{i\alpha } & 0%
\end{array}%
\right) .
\end{eqnarray}%
Other relevant elements in the mapping relation are $U_{\mathbf{1},mj}=0$
and $U_{\mathbf{1},jn}=0$ for $j\neq m,n$; $U_{\mathcal{I},mj}=0$ and $U_{%
\mathcal{I},jn}=0$ for $j\neq m,n$.

The scattering center with the even-parity $C$ symmetry satisfies $%
H_{c}=cH_{c}^{T}c^{-1}$. From the definition $\Delta =H_{c}-\left( 2J\cos
k\right) \mathbf{1}$, we have the relation $\Delta =c\Delta ^{T}c^{-1}$;
therefore, we obtain%
\begin{equation}
\Delta ^{-1}=c\left( \Delta ^{-1}\right) ^{T}c^{-1}.
\end{equation}%
For the $C_{\mathbf{1}}$ symmetry, we have%
\begin{equation}
\Delta _{nm}^{-1}=e^{i\alpha }\Delta _{mn}^{-1},
\end{equation}%
thus, we obtain the symmetric transmission%
\begin{equation}
\left\vert t_{L}\right\vert =\left\vert t_{R}\right\vert .
\end{equation}%
For the $C_{\mathcal{I}}$ symmetry, we have%
\begin{equation}
\Delta _{mm}^{-1}=\Delta _{nn}^{-1},
\end{equation}%
besides, the $C_{\mathcal{I}}$ symmetry also demands $e^{i\alpha }=1$ or $%
\Delta _{mn}^{-1}=\Delta _{nm}^{-1}=0$. Thus, in general case we only have
the symmetric reflection
\begin{equation}
r_{L}=r_{R}.
\end{equation}

The scattering center with the even-parity $K$ symmetry satisfies $%
H_{c}=\Bbbk H_{c}^{\ast }\Bbbk ^{-1}$. Therefore, we obtain the relation $%
\Delta =\Bbbk \Delta ^{\ast }\Bbbk ^{-1}$; and consequently,
\begin{equation}
\Delta ^{-1}=\Bbbk \left( \Delta ^{-1}\right) ^{\ast }\Bbbk ^{-1}.
\end{equation}%
For the $K_{\mathbf{1}}$ symmetry, we obtain%
\begin{equation}
\Delta _{mm}^{-1}=(\Delta _{mm}^{-1})^{\ast },\Delta _{nn}^{-1}=(\Delta
_{nn}^{-1})^{\ast };\Delta _{nm}^{-1}=e^{i\alpha }\left( \Delta
_{nm}^{-1}\right) ^{\ast },\Delta _{mn}^{-1}=e^{-i\alpha }(\Delta
_{mn}^{-1})^{\ast },
\end{equation}%
thus, $\Delta _{mm}^{-1}$, $\Delta _{nn}^{-1}$, and $\Delta _{mn}^{-1}\Delta
_{nm}^{-1}$ are all real numbers; and we have symmetric reflection%
\begin{equation}
\left\vert r_{L}\right\vert =\left\vert r_{R}\right\vert .
\end{equation}%
For the $K_{\mathcal{I}}$ symmetry, we obtain%
\begin{equation}
\Delta _{nm}^{-1}=e^{i\alpha }\left( \Delta _{mn}^{-1}\right) ^{\ast },
\end{equation}%
thus, we have symmetric transmission%
\begin{equation}
\left\vert t_{L}\right\vert =\left\vert t_{R}\right\vert .
\end{equation}%
we also have $\Delta _{mm}^{-1}=\left( \Delta _{nn}^{-1}\right) ^{\ast }$,
which does not lead to a symmetric relation on the scattering coefficients.

The scattering center with the even-parity $Q$ symmetry satisfies $%
H_{c}=qH_{c}^{\dag }q^{-1}$. Therefore, we obtain $\Delta =q\Delta ^{\dagger
}q^{-1}$; and consequently%
\begin{equation}
\Delta ^{-1}=q\left( \Delta ^{-1}\right) ^{\dagger }q^{-1}.
\end{equation}%
For the $Q_{\mathbf{1}}$ symmetry, we have%
\begin{equation}
\Delta _{mm}^{-1}=(\Delta _{mm}^{-1})^{\ast },\Delta _{nn}^{-1}=(\Delta
_{nn}^{-1})^{\ast };\Delta _{nm}^{-1}=e^{i\alpha }\left( \Delta
_{mn}^{-1}\right) ^{\ast },\Delta _{mn}^{-1}=e^{-i\alpha }(\Delta
_{nm}^{-1})^{\ast },
\end{equation}%
the $Q_{\mathbf{1}}$ symmetry also requires $e^{2i\alpha }=1$. From $\Delta
_{nm}^{-1}=e^{i\alpha }\left( \Delta _{mn}^{-1}\right) ^{\ast }$, we obtain $%
\left\vert t_{L}\right\vert =\left\vert t_{R}\right\vert $. Notice that $%
\Delta _{mm}^{-1}$, $\Delta _{nn}^{-1}$, and $\Delta _{mn}^{-1}\Delta
_{nm}^{-1}$ are all real numbers; thus, we have both the symmetric
transmission and reflection%
\begin{equation}
\left\vert t_{L}\right\vert =\left\vert t_{R}\right\vert ,\left\vert
r_{L}\right\vert =\left\vert r_{R}\right\vert .
\end{equation}%
For the $Q_{\mathcal{I}}$ symmetry, we have%
\begin{equation}
\Delta _{mm}^{-1}=\left( \Delta _{nn}^{-1}\right) ^{\ast },\Delta
_{nm}^{-1}=e^{i\alpha }\left( \Delta _{nm}^{-1}\right) ^{\ast },\Delta
_{mn}^{-1}=e^{-i\alpha }\left( \Delta _{mn}^{-1}\right) ^{\ast }.
\end{equation}%
These constraints are insufficient to result in the symmetric transmission
as well as the symmetric reflection.

The scattering center with the even-parity $P$ symmetry satisfies $%
H_{c}=pH_{c}p^{-1}$; therefore, we obtain $\Delta =p\Delta p^{-1}$ and
\begin{equation}
\Delta ^{-1}=p\left( \Delta ^{-1}\right) p^{-1}.
\end{equation}%
For the $P_{\mathbf{1}}$ symmetry, we have
\begin{equation}
\Delta _{nm}^{-1}=e^{i\alpha }\Delta _{nm}^{-1},\Delta
_{mn}^{-1}=e^{-i\alpha }\Delta _{mn}^{-1}.
\end{equation}%
These constraints are insufficient to result in the symmetric transmission
as well as the symmetric reflection.\newline
For the $P_{\mathcal{I}}$ symmetry, we have%
\begin{equation}
\Delta _{mm}^{-1}=\Delta _{nn}^{-1},\Delta _{nm}^{-1}=e^{i\alpha }\Delta
_{mn}^{-1}.
\end{equation}%
Thus, we have both the symmetric transmission and reflection%
\begin{equation}
\left\vert t_{L}\right\vert =\left\vert t_{R}\right\vert ,r_{L}=r_{R}.
\end{equation}%
These conclusions are summarized in Supplemental Table \ref{TableI} and give
the symmetric constraints on the scattering coefficients listed in Table I
of the main text.

\begin{table}[thb]
\caption{Symmetry-protected constraints on the transmission and reflection
coefficients.} \label{TableI}%
\begin{tabular}{|c|c|c|c|c|c|c|c|}
\hline
\textrm{Symmetry} & $C_{\mathbf{1}}$ & $C_{\mathcal{I}}$ & $K_{\mathbf{1}}$
& $K_{\mathcal{I}}$ & $Q_{\mathbf{1}}$ & $P_{\mathcal{I}}$ & $Q_{\mathcal{I}%
},P_{\mathbf{1}}$ \\ \hline
\textrm{Constraint} & $\left\vert t_{L}\right\vert =\left\vert
t_{R}\right\vert $ & $r_{L}=r_{R}$ & $\left\vert r_{L}\right\vert
=\left\vert r_{R}\right\vert $ & $\left\vert t_{L}\right\vert =\left\vert
t_{R}\right\vert $ & $\left\vert t_{L}\right\vert =\left\vert
t_{R}\right\vert ,\left\vert r_{L}\right\vert =\left\vert r_{R}\right\vert $
& $\left\vert t_{L}\right\vert =\left\vert t_{R}\right\vert ,r_{L}=r_{R}$ &
\textrm{None} \\ \hline
\end{tabular}
\end{table}

These conclusions are straightforwardly applicable to explain the many
intriguing symmetric/asymmetric scattering phenomena reported in the
literatures. For example, the $C_{\mathbf{1}}$\ symmetry protected symmetric
transmission \cite{Potton,LJinPRA18}; the $C_{\mathcal{I}}$\ symmetry
protected symmetric reflection \cite%
{XZZhang13,CLi17,LXQ15,LJin16,LJinPRL,BYan,Longhi15}; the $K_{\mathbf{1}}$
symmetry protected symmetric reflection \cite{Cannata,LJinSR}; the $K_{%
\mathcal{I}}$ symmetry protected symmetric transmission \cite%
{LJinSR,Muga,LonghiJRA11,Longhi10,Stone11,Lin11,Feng11,Kottos,LGe12,Regensburger12, Feng13,Engheta13,Schomerus,Ahmed,Ambichl13,Savoia,Mostafazadeh,Wu14,Ramezani14,Fleury15, JLi15,YHuang15,Miri16,HZhao16,JHWu,LGe16,Wong16,LuPRL16,Ramezani16,JLi16,Ali16,JLi17,JHWuPRA17,LGe17,Schmelcher17,Gong17,Jin18,YLai18,YZhang18,Koutserimpas18,ShenPRM18,LeeOE18,Christensen18,Zhi18,Wu19,Muga19,Zhao19,SweeneyPRL19,Sweeney19,Novitsky20,Yang20}%
; the $Q_{\mathbf{1}}$ symmetry protected symmetric transmission and
reflection \cite{LJin12,Sounas}; and the $P_{\mathcal{I}}$ symmetry
protected symmetric transmission and reflection \cite%
{Chong10,Wan2011,CPA,Baranov17,LonghiOL18,Trainiti19,Zhong20,Haque20}.
Without the protection of these symmetries, both the transmission and
reflection are asymmetric \cite{LJinNJP17,LJinPRL,LDu20}.

\subsection*{B: Details for the light scattering in the three-coupled-resonator
scattering center}

The scattering center of the three coupled resonators is schematically
illustrated in the main text Fig.~2(a). The scattering center encloses a
synthetic magnetic flux $\phi $ induced by the Peierls phase factor in the
coupling. We now analyze the properties of the scattering center through the
two-port scattering dynamics. The leads $1$ and $3$ are under consideration;
the lead $2$ is disconnected from the scattering center.

For the case $\left\{ V_{1},V_{2},V_{3}\right\} =\left\{ i\gamma ,-i\gamma
,0\right\} $, the Hamiltonian $H_{c}$ is not symmetry-protected. The
scattering coefficients are obtained in the form of%
\begin{eqnarray}
t_{L} &=&\frac{J(e^{2ik}-1)(Je^{i\phi }+i\gamma +2J\cos k)}{%
(2-e^{-2ik}+e^{2ik}+e^{i(k-\phi )}+e^{i(k+\phi )})J^{2}+iJ\gamma
e^{ik}-\gamma ^{2}}, \\
r_{L} &=&-\frac{(1+e^{2ik}+e^{i(k-\phi )}+e^{i(k+\phi )})J^{2}+iJ\gamma
e^{-ik}-\gamma ^{2}}{(2-e^{-2ik}+e^{2ik}+e^{i(k-\phi )}+e^{i(k+\phi
)})J^{2}+iJ\gamma e^{ik}-\gamma ^{2}}, \\
t_{R} &=&\frac{Je^{-i\phi }(e^{2ik}-1)(J+i\gamma e^{i\phi }+2Je^{i\phi }\cos
k)}{(2-e^{-2ik}+e^{2ik}+e^{i(k-\phi )}+e^{i(k+\phi )})J^{2}+iJ\gamma
e^{ik}-\gamma ^{2}}, \\
r_{R} &=&-\frac{(1+e^{2ik}+e^{i(k-\phi )}+e^{i(k+\phi )})J^{2}+iJ\gamma
e^{3ik}-\gamma ^{2}e^{2ik}}{(2-e^{-2ik}+e^{2ik}+e^{i(k-\phi )}+e^{i(k+\phi
)})J^{2}+iJ\gamma e^{ik}-\gamma ^{2}}.
\end{eqnarray}%
The scattering coefficients are depicted in Figs.~2(b) and 2(c) of the main
text; and they diverge at $J=\gamma /2$ and $\phi =\pm \pi /2$ for the
resonant inputs at the momentum $k=-\pi /2$. In the general case, the
transmission and reflection are asymmetric; these are reflected from the
four elements of the inverse matrix $\Delta ^{-1}$
\begin{equation}
\left(
\begin{array}{cc}
\Delta _{mm}^{-1} & \Delta _{mn}^{-1} \\
\Delta _{nm}^{-1} & \Delta _{nn}^{-1}%
\end{array}%
\right) =\frac{J^{2}}{\det \Delta }\left(
\begin{array}{cc}
1+e^{2ik}+e^{-2ik}+i\left( \gamma /J\right) \left( e^{ik}+e^{-ik}\right) &
e^{ik}+e^{-ik}+e^{-i\phi }+i\left( \gamma /J\right) \\
e^{ik}+e^{-ik}+e^{i\phi }+i\left( \gamma /J\right) & 1+e^{2ik}+e^{-2ik}+%
\left( \gamma /J\right) ^{2}%
\end{array}%
\right) .
\end{equation}%
where $\Delta =H_{c}-\left( 2J\cos k\right) \mathbf{1}$. The determinant is $%
\det \Delta =J^{3}\left[ e^{-i\phi }+e^{i\phi }-e^{-3ik}-e^{3ik}-\left(
\gamma ^{2}/J^{2}\right) \left( e^{-ik}+e^{ik}\right) \right] $. Then, we
obtain the contrast between the transmission and reflection of the\ inputs
from the forward and backward directions for the resonant incidence with the
momentum $k=-\pi /2$%
\begin{equation}
\frac{t_{R}}{t_{L}}=\frac{e^{-i\phi }+i\gamma /J}{e^{i\phi }+i\gamma /J},%
\frac{r_{R}}{r_{L}}=\frac{2\cos \phi -i\gamma /J+i\gamma ^{2}/J^{2}}{2\cos
\phi -i\gamma /J-i\gamma ^{2}/J^{2}}.
\end{equation}%
At $\gamma =J$, the special cases of $\phi =-\pi /2$ for a unidirectional
transmissionless $t_{L}=0$, $r_{L}=1$; $t_{R}=-2i$, $r_{R}=0$ and $\phi =\pi
/2$ for a unidirectional absorption $t_{L}=-2i$, $r_{L}=1$; $t_{R}=0$, $%
r_{R}=0$ are shown in the main text Fig. 3.

\subsection*{C: Non-Hermitian three-coupled-resonator scattering center with
dissipative coupling}

In the coupled mode theory, the equations of motion for the three-site
scattering center with dissipative coupling $-i\kappa $ are given by%
\begin{eqnarray}
i\dot{\phi}_{c}^{k}(1) &=&\omega _{0}\phi _{c}^{k}(1)-i\kappa \phi
_{c}^{k}(2)+J\phi _{c}^{k}(3)+J\phi _{1}^{k}\left( -1\right) , \\
i\dot{\phi}_{c}^{k}(2) &=&\omega _{0}\phi _{c}^{k}(2)-i\kappa \phi
_{c}^{k}(1)+Je^{-i\phi }\phi _{c}^{k}(3), \\
i\dot{\phi}_{c}^{k}(3) &=&\omega _{0}\phi _{c}^{k}(3)+J\phi
_{c}^{k}(1)+Je^{i\phi }\phi _{c}^{k}(2)+J\phi _{3}^{k}\left( 1\right) ,
\end{eqnarray}%
where $\phi _{1}^{k}\left( -1\right) $ and $\phi _{3}^{k}\left( 1\right) $
are the wavefunctions of the sites on the leads $1$ and $3$ that are
connected with the scattering center sites $1$ and $3$, respectively. The
dissipative coupling is reciprocal and is directly induced by the
dissipation in the link resonator between the primary resonators $1$ and $2$
\cite{HCWu20}. The link resonator and the primary resonators are on
resonant. The effective dissipative coupling strength $\kappa =\kappa
_{0}^{2}/\gamma $ is inversely proportional to the link resonator
dissipation $\gamma $ and quadratically proportional to the coupling
strength between the link resonator and the primary resonators $\kappa _{0}$
\cite{LYou,CHHu20}. The dissipative coupling has also been experimentally
realized in many anti-parity-time symmetric systems \cite%
{antiPT,Konotop,CWQiu,SWDu}.

The scattering center is described by the Hamiltonian%
\begin{equation}
H_{c}=\left(
\begin{array}{ccc}
0 & -i\kappa & J \\
-i\kappa & 0 & Je^{-i\phi } \\
J & Je^{i\phi } & 0%
\end{array}%
\right) ,
\end{equation}%
the scattering properties of which are not symmetry-protected.

The contrast between the transmission and reflection from opposite incident
directions are calculated as follows. We obtain the four elements of the
inverse matrix $\Delta ^{-1}$ in the form of%
\begin{equation}
\left(
\begin{array}{cc}
\Delta _{mm}^{-1} & \Delta _{mn}^{-1} \\
\Delta _{nm}^{-1} & \Delta _{nn}^{-1}%
\end{array}%
\right) =\frac{J^{2}}{\det \Delta }\left(
\begin{array}{cc}
e^{2ik}+1+e^{-2ik} & e^{ik}+e^{-ik}-i\left( \kappa /J\right) e^{-i\phi } \\
e^{ik}+e^{-ik}-i\left( \kappa /J\right) e^{i\phi } & e^{2ik}+2+e^{-2ik}+%
\left( \kappa /J\right) ^{2}%
\end{array}%
\right) .
\end{equation}%
In general case, the transmission and reflection are asymmetric.

However, for the resonant incidence with the momentum $k=-\pi /2$, the
transmission is symmetric $\left\vert t_{L}\right\vert =\left\vert
t_{R}\right\vert $ because of $\left\vert \Delta _{nm}^{-1}\right\vert
=\left\vert \Delta _{mn}^{-1}\right\vert $; but the reflection is
asymmetric. To break the reciprocity at $k=-\pi /2$, the on-site terms $%
\left\{ V_{1},V_{2},V_{3}\right\} $ of the scattering center are helpful%
\begin{equation}
H_{c}=\left(
\begin{array}{ccc}
V_{1} & -i\kappa & J \\
-i\kappa & V_{2} & Je^{-i\phi } \\
J & Je^{i\phi } & V_{3}%
\end{array}%
\right) .  \label{Hc_dis}
\end{equation}%
For the resonant incidence $k=-\pi /2$, the four elements of the inverse
matrix $\Delta ^{-1}$ are in the form of%
\begin{equation}
\left(
\begin{array}{cc}
\Delta _{mm}^{-1} & \Delta _{mn}^{-1} \\
\Delta _{nm}^{-1} & \Delta _{nn}^{-1}%
\end{array}%
\right) =\frac{1}{\det \Delta }\left(
\begin{array}{cc}
V_{2}V_{3}-J^{2} & -JV_{2}-iJ\kappa e^{-i\phi } \\
-JV_{2}-iJ\kappa e^{i\phi } & V_{1}V_{2}+\kappa ^{2}%
\end{array}%
\right) ,
\end{equation}%
where the determinant $D\equiv \det \Delta
=V_{1}V_{2}V_{3}-J^{2}V_{1}-J^{2}V_{2}+\kappa ^{2}V_{3}-iJ^{2}\kappa \left(
e^{-i\phi }+e^{i\phi }\right) $.

At $\phi =\pm \pi /2$, the transmission is asymmetric for the real $V_{2}$;
the transmission is symmetric for the imaginary $V_{2}$. For the real $%
V_{1,2,3}$, the reflection is symmetric; otherwise, the reflection is
asymmetric.

We take $V_{1}=V_{3}=0$ as an illustration. The scattering coefficients for
the momentum $k=-\pi /2$ are obtained in the form of%
\begin{eqnarray}
t_{L} &=&\frac{J(e^{2ik}-1)(-iV_{2}+\kappa e^{i\phi }+2iJ\cos k)}{%
iJ^{2}e^{-2ik}(e^{4ik}+e^{2ik}-1)+J(iV_{2}e^{-ik}-iV_{2}e^{ik}+\kappa
e^{i(k-\phi )}+\kappa e^{i(k+\phi )})-i\kappa ^{2}}, \\
r_{L} &=&-\frac{e^{-i\phi }(e^{i(k+\phi )}J-i\kappa )(iJe^{ik}+\kappa
e^{i\phi })}{iJ^{2}e^{-2ik}(e^{4ik}+e^{2ik}-1)+J(iV_{2}e^{-ik}-iV_{2}e^{ik}+%
\kappa e^{i(k-\phi )}+\kappa e^{i(k+\phi )})-i\kappa ^{2}}, \\
t_{R} &=&\frac{iJe^{-i\phi }(e^{2ik}-1)(-V_{2}e^{i\phi }-i\kappa +2Je^{i\phi
}\cos k)}{iJ^{2}e^{-2ik}(e^{4ik}+e^{2ik}-1)+J(iV_{2}e^{-ik}-iV_{2}e^{ik}+%
\kappa e^{i(k-\phi )}+\kappa e^{i(k+\phi )})-i\kappa ^{2}}, \\
r_{R} &=&-\frac{e^{-i\phi }(e^{i(k+\phi )}\kappa +iJ)(Je^{i\phi }-i\kappa
e^{ik})}{iJ^{2}e^{-2ik}(e^{4ik}+e^{2ik}-1)+J(iV_{2}e^{-ik}-iV_{2}e^{ik}+%
\kappa e^{i(k-\phi )}+\kappa e^{i(k+\phi )})-i\kappa ^{2}}.
\end{eqnarray}%
Both the scattering coefficients $t$ and $r$ diverge at $\phi =\pm \arccos
\left( J^{2}-\kappa ^{2}\right) \left( 2J\kappa \right) ^{-1}$ and $V_{2}=0$
when the resonant momentum $k=-\pi /2$. For the real $V_{2}=J$, we have the
scattering coefficients $t_{L}=-2i,r_{L}=-i;t_{R}=0,r_{R}=i$ at $\phi =-\pi
/2$; and we have the scattering coefficients $%
t_{L}=0,r_{L}=-i;t_{R}=-2i,r_{R}=i$ at $\phi =\pi /2$. The numerical
simulations are shown in Fig.~\ref{figS1}.

\begin{figure}[thb]
\includegraphics[ bb=0 0 590 135, width=17.8 cm, clip]{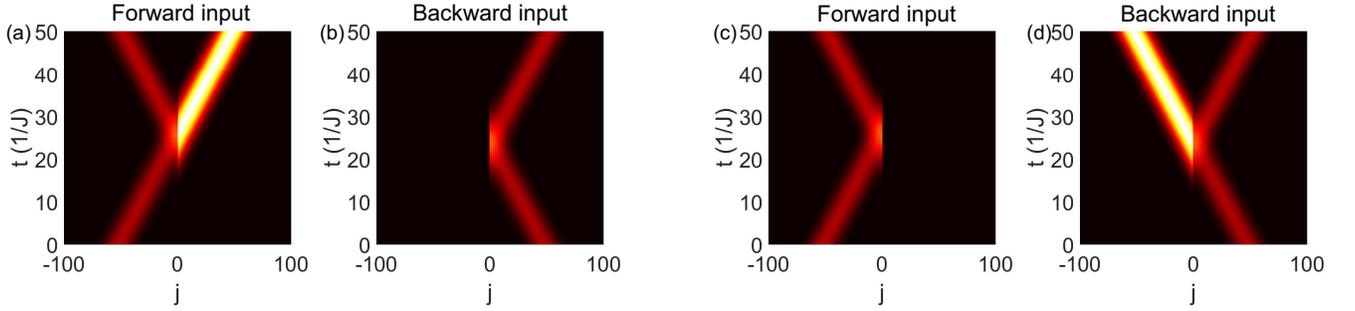}
\caption{Numerical simulation of the asymmetric transmission in the
three-site center with dissipative coupling [Eq.~(\protect\ref{Hc_dis})].  The incidences are Gaussian wave packet with the momentum $k=-\pi/2$ in the simulations. The excitation intensity is depicted. (a) Forward and (b) backward incidence for $\protect\phi =-\protect\pi /2$. $\left\vert t_{L}\right\vert ^{2}=4$, $\left\vert r_{L}\right\vert ^{2}=1$; $\left\vert t_{R}\right\vert ^{2}=0$, $\left\vert r_{R}\right\vert ^{2}=1$.
(c) Forward and (d) backward incidence for $\protect\phi =\protect\pi /2$. $\left\vert t_{L}\right\vert ^{2}=0$, $\left\vert r_{L}\right\vert ^{2}=1$; $\left\vert t_{R}\right\vert ^{2}=4$, $\left\vert r_{R}\right\vert ^{2}=1$.
The dissipative coupling is $\protect\kappa =J=1$ and the on-site terms are $\{V_{1},V_{2},V_{3}\}=\{0,J,0\}$. The incident frequency is $\protect\omega_0$.}
\label{figS1}
\end{figure}

\clearpage
\end{widetext}


\begin{thebibliography}{999}
\bibitem{Diff} We emphasize two aspects in the definitions of symmetries
that distinct from the topological classification~\cite{KawabataPRX19}: (i)
the signs $\epsilon _{\Bbbk }$ and $\epsilon _{q}$ in the $K$ and $Q$
symmetries; which are absent in the definitions of the $K$ and $Q$
symmetries for topological classification because $H_{c}$ and $iH_{c}$ have
identical topological properties. However, the scattering properties of $%
H_{c}$ and $iH_{c}$ are not identical. The symmetries with opposite parities
impose different scattering features; (ii) the even-parity $P$ symmetry.
\bibitem{E8} The combination of two symmetries that belong to the same type
gives the $P$ symmetry. The combination of the $P$ symmetry and any of the $%
C,K,Q$ symmetries does not alter the $C,K,Q$ symmetries. The combination of
any two of the three symmetries $C,K,Q$ produces the third symmetry.
\bibitem{NoteInverseMatrix} If the determint $\left\vert \Delta \right\vert $
of matrix $\Delta $ vanishes, just regarding the inverse matrix $\Delta
^{-1} $ as the adjoint matrix $\Delta ^{\mathrm{adj}}$ of matrix $\Delta $
in Eqs. (\ref{T}) and (\ref{R}). They satisfies the relation $\Delta
^{-1}=\Delta ^{\mathrm{adj}}/\left\vert \Delta \right\vert $.
\end{thebibliography}
\end{document}